\documentclass[a4paper,12pt]{article}
\usepackage{latexsym}
\usepackage[dvips]{graphicx}
\usepackage{amssymb}
\usepackage{graphicx}

            \newcommand{\be}{\begin{equation}}
            \newcommand{\ee}{\end{equation}}
            \newcommand{\bearray}{\begin{eqnarray}}
            \newcommand{\eearray}{\end{eqnarray}}
            \newcommand{\bee}[1]{\begin{equation}\label{#1}}
            \newcommand{\bey}{\begin{eqnarray}}
            \newcommand{\byy}[1]{\begin{eqnarray}\label{#1}}
            \newcommand{\eey}{\end{eqnarray}}
            \newcommand{\R}[1]{(\ref{#1})}
            \newcommand{\C}[1]{\cite{#1}}

            \newcommand{\mvec}[1]{\mbox{\boldmath{$#1$}}}

            \newcommand{\x}{(\mvec{x},t)}

            \newcommand{\bx}{\mvec{\Box}}

\begin{document}

\title{Comment on:\\I-Shih Liu:
Constitutive theory\\of anisotropic rigid heat conductors}
\author{W. Muschik\thanks{muschik@physik.tu-berlin.de}\  \\
Institut f\"{u}r Theoretische Physik\\Technische Universit\"{a}t Berlin\\
Hardenbergstra$\ss$e 36\\D-10623 Berlin, Germany\\}
\date{\today}
\maketitle
\abstract\noindent
In I-Shih Liu's paper \C{1}, the compatibility of anisotropy and 
material frame 
indifference of a rigid heat conductor is investigated. For this purpose, the 
deformation gradient is introduced into the domain of the constitutive
mapping. Because of the presupposed rigidity, the deformation gradient
is here represented by an 
orthogonal tensor. The statement, that the usual procedure -- not to introduce
the deformation gradient into the state space of rigid heat conductors --
causes isotropy because of the material frame indifference, is misleading.

\section{Introduction}

In \C{1}, the following material axioms are postulated
\begin{itemize}
\item rigid body motion as internal mechanical constraint,
\item principle of material frame indifference,
\item material symmetry,
\item entropy principle.
\end{itemize}
Because we not interested in the entropy production, the entropy principle is
not considered here. The three other principles are used in
formulations and interpretations which are slightly different from the 
usual ones. As in \C{1}, only acceleration-insensitive materials are 
considered \C{2}.

The main point is that material frame indifference and material symmetry
belong to different properties of the material mapping which are independent
of each other, also for rigid material, as we will see later on.

Material symmetry is an observer-independent property of the material mapping
which therefore is described by a tensor equation being valid for all 
observers. 
In contrast to material symmetry, material frame indifference is a property
of the components of the material mapping which belong to different observers
\C{3}.
Changing the observers results in changing the components of the material
mapping, whereas the material mapping itself remains observer independent.
Consequently, we have strictly to distinguish between the material mapping
itself and its components, because the homomorphism between tensors and their
components is not true with respect to material frame indifference. If this 
fact is not taken into account, the strange statement results ``all 
rigid materials are isotropic''\footnote{a statement which is cited from 
Liu's introduction}. Consequently, Liu (and also we) ``shall correctly 
formulate a constitutive theory of rigid heat conductors (\C{1})'',or
rather of rigid materials.
But we will do that in an other way as Liu: He introduces the non-objective
rotational part of the deformation gradient into the state space, whereas we
formulate the principle of material indifference correctly in 
observer-dependent components of the observer-independent material mapping.  

The paper is organized as follows: After introducing the material mapping,
rigidity, material symmetry, material frame indifference and isotropy and
their interdependence are discussed. Finally the simple example of Fourier
heat conduction in a rigid heat conductor is considered for elucidation.

\section{The material mapping}

Material is described by a constitutive mapping ${\cal M}$ 
\bee{0}
{\cal M}:\ \mathbb{Z}\ \rightarrow\ \mathbb{M},\quad\mvec{z}\ \mapsto\
\mvec{M} 
\ee
whose domain $\mathbb{Z}$ is spanned by the state space and its range 
$\mathbb{M}$ by the constitutive properties.
For the example of Fourier heat conduction, the gradient of
temperature is in the
state space and the heat flux in the space of the constitutive 
properties (often, the constitutive mapping itself is denoted as heat flux).

We presuppose that the elements of the
state space $\mvec{z}\x$ and of the constitutive properties $\mvec{M}\x$  are 
Euclidean tensors: they are ``objective'' (a presupposition which is not really
necessary \C{3}). Because in general, materials are acceleration-sensitive,
we need a so-called ``2nd entry'' $\Box$ in the domain of the constitutive 
mapping, which describes the accelerated motion of the material with respect
to a freely chosen standard frame of reference \C{4}:  
\bee{1}
\mvec{M}\x\ =\ {\cal M}(\mvec{z}\x , \bx ).
\ee
Because here --as in \C{1}-- only acceleration-insensitive materials 
are considered, we obtain from \R{1} by ignoring the 2nd entry
\bee{2}
\mvec{M}\x\ =\ {\cal M}(\mvec{z}\x ).
\ee

For the special considerations in the sequel, we restrict ourselves to
material mappings whose domain and range are presupposed to be spanned
by Euclidean tensors of first order\footnote{the general case is treated in 
\C{3}}.

We now consider two observers, $\mvec{B}$ and $\mvec{B}^*$, with their bases
$\{\mvec{e}_j\}$ and $\{\mvec{e}_j^*\}$. Consequently, the component 
representations of the constitutive equation \R{2} are
\byy{3}
M_j\ =\ \mvec{e}_j \cdot {\cal M}(z_m\mvec{e}_m),&&\qquad
M_j^*\ =\ \mvec{e}_j^* \cdot {\cal M}(z_m^*\mvec{e}_m^*), \\ \label{4}
M_j\ =\ {\cal M}_j(z_m),\hspace{.93cm}&&\qquad
M_j^*\ =\ {\cal M}_j^*(z_m^*).
\eey

Here, the transfer from \R{2} to \R{3} and \R{4}, respectively, is made by
using a part of the principle of material frame indifference:  
\begin{itemize}
\item the material mapping ${\cal M}$ in \R{2} is observer-independent
(no $^*$ at ${\cal M}$ in \R{3}$_2$).
\end{itemize}
But as can be seen from \R{4}, this statement is not true for the component
representation. Different observers see different components of ${\cal M}$:
${\cal M}_j$ and ${\cal M}_j^*$.

\section{Rigid material}

Because later on rigid heat conductors are considered, according to \C{1}, 
we have to choose a to rigidity adapted deformation gradient
\bee{01}
\mvec{F}\ =\ \mvec{R}\cdot\mvec{U},\quad \mbox{with } \ \mvec{U}\ \doteq\
\mvec{1},\qquad\mvec{R}\in {\cal O}.
\ee
Thus in a rigid materials, the deformation is ``switched off'', 
the Cauchy-Green tensors are unit tensors, and --as in \C{1}-- the
deformation gradient includes only its rotational part. Because the  
Cauchy-Green tensors are constant, they do not appear in the state
space $\mathbb{Z}$. Also $R$ is not an element of the state space,
because it belongs to an inaccelerated motion of the material\footnote{ 
the relative
rotation between the referential and the current frame which is independent 
of time. 
Otherwise the material is accelerated and the 2nd entry in
\R{1} cannot be ignored} which do not influence
the material properties. Consequently, we will not accept I-Shih Liu's
eqs.(1) and (2), where $\mvec{F}$, or better $\mvec{R}$, appears in
the state space without any influence on the constitutive properties. 
By taking these reasons into account, we will start 
out later on with an example using the ``simple-minded formulation \C{1}''
of the heat flux in rigid heat conductors
\bee{02}
\mvec{q}\ =\ {\cal W}(\Theta,\nabla\Theta)
\ee
without introducing the non-objective rotational part $\mvec{R}$ of the 
deformation gradient into the state space of that rigid material. If the state
space does not contain any of the Cauchy-Green tensors, the material is
necessarily rigid, because a deformation cannot be described by the chosen 
state space.

\section{The material symmetry}

Presupposing a symmetry group ${\cal G}$ whose elements $\mvec{H}$ are 
tensors of second order defined on the current configuration space, 
the material property $\mvec{M}$, element of the range of the 
constitutive mapping, can be represented by different group 
generated constitutive mappings
\bee{5}
\mvec{M}\ =\ {\cal M}^{\mvec{H}}(\mvec{z}\x)\ =\ 
{\cal M}^{\mvec{1}}(\mvec{z}\x)\ \equiv\ {\cal M}(\mvec{z}\x),\quad
\wedge \mvec{H}\in {\cal G}.
\ee
\begin{itemize}
\item The symmetry axiom is:
\end{itemize}
\bee{6}\wedge \mvec{H}\in {\cal G}:\quad
{\cal M}^{\mvec{H}}(\mvec{z}\x)\ \doteq\ \mvec{H}^{-1}\cdot
{\cal M}(\mvec{H}\cdot\mvec{z}\x)\ =\ \mvec{M}\ =\ {\cal M}(\mvec{z}\x).
\ee
The last two equations follow from \R{5}. The symmetry axiom, consisting of
\R{5} and \R{6}, corresponds to eq.(4) in \C{1}. Thus, the material mapping
satifies eq.(6) in \C{1} without use of the roundabout way of introducing 
$\mvec{R}$ into the state space with the aim of eliminating it later on 
\bee{7}
\mvec{H}^{-1}\cdot{\cal M}(\mvec{H}\cdot\mvec{z}\x)\ =\ {\cal M}(\mvec{z}\x).
\ee
This equation is generated by symmetry properties of the material and has no
connection to the principle of material frame indifference which is
discussed now.

\section{Material frame indifference}

We now refer to the two already introduced observers $\mvec{B}$ and 
$\mvec{B}^*$
in \R{4}, and we write down their component equations of \R{7} 
\bee{8}
M_j\ =\ H^{-1}_{jm}{\cal M}_m (H_{pq}z_q),\qquad
M_j^*\ =\ H^{*-1}_{jm}{\cal M}_m^* (H^*_{pq}z_q^*).
\ee
Here we obtain a further part of the principle of material frame indifference:
\begin{itemize}
\item the constitutive equations written in components belonging to
the observers are observer-invariant (frame- or form-invariant).
\end{itemize}

Because we presuppose that
\begin{itemize}
\item domain and range of the constitutive mapping are spanned by objective
quantities, that means, these quantities are Euclidean tensors,
\end{itemize}
and consequently, we obtain
\begin{itemize}
\item changing the observer is performed by an orthogonal transformation of
their tensor components:
\end{itemize}
\bee{9}Q_{ik}\in{\cal O}:\qquad
M_j^*\ =\ Q_{jp}M_p,\quad H^*_{pq}\ =\ Q_{pr}H_{rs}Q^{-1}_{sq}, \quad
z^*_q\ =\ Q_{qn}z_n.
\ee
Consequently, \R{8}$_2$ results by use of \R{8}$_1$ in
\byy{10}
Q_{ja}M_a &=& Q_{jb}H^{-1}_{bc}Q^{-1}_{cm}{\cal M}_m^* 
(Q_{pq}H_{qr}Q_{rm}^{-1}Q_{ms}z_s),\\ \label{11}
M_j &=& H^{-1}_{jc}Q^{-1}_{cm}{\cal M}_m^* 
(Q_{pq}H_{qr}z_r)\ =\ H^{-1}_{jm}{\cal M}_m (H_{pq}z_q) .
\eey
Consequently, we obtain the connection of the components of the material
mapping belonging to different observers by taking rigidity (no
deformation gradient in he state space),
anisotropy and material frame indifference into account
\bee{12}
Q^{-1}_{cm}{\cal M}_m^*(Q_{pq}H_{qr}z_r)\ =\ {\cal M}_c (H_{pq}z_q). 
\ee
This equation replaces eq.(1) in \C{1} which we do not accept for two
reasons. The first one was already mentioned: It contains the relative
rotation $\mvec{R}$ which has no influence on the material properties
and secondly, it is not related to any changing the observers
(frames). To our insight, material frame indifference means \C{3}: The
material mapping ${\cal M}$ is observer-independent, and how do
transform the components of the material mapping belonging to the
observers by changing them. This tranformation is given by \R{12}, if
a symmetry group is present. 

Taking the identity of the group, \R{12} becomes
\bee{13}
\mvec{H}\doteq\mvec{1}\ \rightarrow\ H_{pq}=\delta_{pq}\ \rightarrow\
Q^{-1}_{cm}{\cal M}_m^*(Q_{pq}z_q)\ =\ {\cal M}_c (z_p). 
\ee
This equation does not represent an isotropic function neither for the 
components of the material mapping nor for the material mapping itself.
This fact is clear, because changing the observer by \R{9} has nothing 
to do with symmetry properties of the material which are described by
\R{7} or for arbitrary observers by \R{8}.

\section{Isotropy}

If the material is isotropic, we obtain from \R{7} the material mapping as an 
isotropic function
\bee{14}
\mvec{H}\doteq\mvec{R},\ \ \mvec{R}\in{\cal O},\qquad 
\mvec{R}^{-1}\cdot{\cal M}(\mvec{R}\cdot\mvec{z}\x)\ =\ {\cal M}(\mvec{z}\x).
\ee
According to \R{8}, the isotropy of the material is valid for each observer
\bee{15}
{\cal M}_j(z_p)\ =\ R^{-1}_{jm}{\cal M}_m (R_{pq}z_q),\qquad
{\cal M}_j^*(z^*_p)\ =\ R^{*-1}_{jm}{\cal M}_m^* (R^*_{pq}z_q^*).
\ee
In this isotropic case, \R{12} results in
\bee{16}
Q^{-1}_{cm}{\cal M}_m^*(Q_{pq}R_{qr}z_r)\ =\ {\cal M}_c (R_{pq}z_q),
\ee
which in accordance with \R{13} for $\mvec{R}\doteq\mvec{1}$. Inserting
\R{16} into \R{15}$_1$ results in
\bee{16a}
{\cal M}_j(z_p)\ =\ R^{-1}_{jm}Q^{-1}_{mn}{\cal M}_n^*(Q_{pq}R_{qr}z_r).
\ee

Both the observers --$\mvec{B}$ and $\mvec{B}^*$-- are arbitrarily, but now
fixed chosen. Consequently, the $Q_{ik}$ in \R{9}$_1$ are fixed. The isotropic
material now undergoes a special unaccelerated motion which is defined by
\bee{16b}
Q_{pq}R_{qr}\ \doteq\ \delta_{pr}\quad\rightarrow\quad
R_{qr}\ =\ Q^{-1}_{qr}.
\ee
The special choice transfers \R{16a} into
\bee{b17}
{\cal M}_j(z_p)\ =\ {\cal M}_j^* (z_p)\quad \rightarrow\quad
{\cal M}_c(\bullet)\ =\ {\cal M}_c^*(\bullet),
\ee
that means: the components of the material mapping are observer
independent, if the material is isotropic.

For exploiting \R{15}$_2$, we need $R^*_{pq}$. According to the transformation
property \R{9}$_3$ of tensors of second order by changing the observer, we 
obtain by use of \R{16b}$_{1,2}$
\bee{b17a}
R^*_{pq}\ =\ Q_{pr}R_{rs}Q^{-1}_{sq}\ =\ Q^{-1}_{pq}\ =\ R_{pq},
\ee
that means, the special chosen symmetry transformation \R{16b}$_1$ generates
equal components for both the observers
\bee{b17b} 
\mvec{e}^*_p \cdot \mvec{R}\cdot\mvec{e}^*_q\ =\
\mvec{e}_p \cdot \mvec{R}\cdot\mvec{e}_q.
\ee
Inserting \R{b17}$_1$, \R{9}$_4$ and \R{b17a} into \R{15}$_2$, results in
\bee{b17c}
{\cal M}_j(Q_{pn}z_n)\ =\ Q_{jm}{\cal M}_m(Q^{-1}_{pq}Q_{qn}z_n)\
\rightarrow\
\mvec{Q}^{-1}\cdot{\cal M}(\mvec{Q}\cdot\mvec{z}\x)\ =\ {\cal M}(\mvec{z}\x).
\ee
Consequently, we obtain 
an isotropic function for the material mappings of
isotropic materials. Material frame indifference --represented by \R{12} and
\R{13}-- is valid for all materials independent of their symmetry. The 
component equations \R{12} and \R{13} result in \R{b17c}$_2$ only if the 
material is isotropic. 
Presupposing wrongly ``\R{b17c}$_2$ as material frame indifference'',
one sticks to isotropic materials, a fact which is clearly not
correct.

If $\mvec{z}=\mvec{0}$ is in the domain of the material mapping, we obtain
from \R{14}
\bee{18} 
\mvec{R}^{-1}\cdot{\cal M}(\mvec{0})\ =\ {\cal M}(\mvec{0}),\qquad\wedge\
\mvec{R}\in{\cal O}.
\ee
This results in
\bee{19}
{\cal M}(\mvec{0})\ =\ \mvec{0}\ \rightarrow\ 
{\cal M}(\mvec{z}\x)\ =\ \mvec{L}(\mvec{z}\x)\cdot\mvec{z}\x.
\ee
Taking \R{14}$_3$ into account, this yields an isotropic tensor function for
$\mvec{L}$
\bee{20}
\mvec{L}(\mvec{z}\x)\ =\ 
\mvec{R}^{-1}\cdot\mvec{L}(\mvec{R}\cdot\mvec{z}\x)\cdot\mvec{R},\qquad\wedge\
\mvec{R}\in{\cal O}.
\ee
If the material is linear, that means, $\mvec{L}$ does not depend on
the state space, we obtain from \R{20} for linear isotropic material
according to Schur's lemma 1 
\bee{27}
\partial\mvec{L}/\partial\mvec{z}={\bf 0}:\qquad \mvec{L}\ =\
\alpha\mvec{1},\quad\alpha=\mbox{const} .
\ee

\section{An example}

We now consider the simple example of Fourier heat conduction in a
rigid heat conductor, not for developing a new theory --see \C{1}-- but
only for elucidation of the ideas used above.

The material equation for the heat flux density is
\bee{28}
\mvec{q}\x\ =\ {\cal M}(\Theta\x,\nabla\Theta\x),
\ee
with the temperature $\Theta$. Because there is no Cauchy-Green tensor
(and of course no deformation gradient) included in the state space,
the heat conductor is rigid. Because of
\bee{29}
\nabla\Theta\ =\ \mvec{0}\ \rightarrow\ \mvec{q}\ =\ \mvec{0},
\ee
the material mapping has the shape
\bee{30}
\mvec{q}\ =\ \mvec{\kappa}(\Theta,\nabla\Theta)\cdot\nabla\Theta.
\ee
The heat conductor is anisotropic and has the symmetry group
\bee{31}
\wedge\mvec{H}\in{\cal G}:\qquad
\mvec{H}\cdot\mvec{q}\ =\
\mvec{H}\cdot\mvec{\kappa}(\Theta,\nabla\Theta)\cdot
\mvec{H}^{-1}\cdot\mvec{H}\cdot\nabla\Theta.
\ee
According to \R{7}, we have in the spatial configuration
\bee{32}
\mvec{H}\cdot\mvec{\kappa}(\Theta,\nabla\Theta)\ =\
\mvec{\kappa}(\Theta,\mvec{H}\cdot\nabla\Theta)\cdot\mvec{H}.
\ee

For the observers $\mvec{B}$ and $\mvec{B}^*$, \R{30} results in
\bee{33}
q_j\ =\ \kappa_{jk}(\Theta,\partial_{r}\Theta)\partial_{k}\Theta,\qquad
q_j^*\ =\ \kappa_{jk}^*(\Theta^*,\partial_{r}^*\Theta)\partial_{k}^*\Theta,
\ee
Material frame indifference consists of two statements
\byy{34}
1:&&\qquad\mvec{\kappa}(\bullet,\bullet)\ \mbox{ is observer-independent}
\\ \label{35}
2:&&\qquad q_j^*\ =\ Q_{jm}q_m,\quad\Theta^*\ =\ \Theta,\quad
\partial_{k}^*\Theta\ =\ Q_{km}\partial_m\Theta.
\eey
Inserting \R{35} into \R{33} results in an equation which is analogous
to \R{13}
\bee{36}
\wedge Q_{ik}\in{\cal O}:\qquad
\kappa_{mp}(\Theta,\partial_{k}\Theta)\ =\ Q^{-1}_{mj}
\kappa_{jk}^*(\Theta,Q_{rs}\partial_{s}\Theta)Q_{kp}.
\ee
Clear is: material frame indifferenc does not generate isotropy.

For taking isotropy into account, $\mvec{H}\in{\cal G}$ has to be
replaced by $\mvec{R}\in{\cal O}$. Consequently, \R{32} becomes
analogous to \R{20} an isotropic tensor function
\bee{37}
\mvec{R}\cdot\mvec{\kappa}(\Theta,\nabla\Theta)\ =\
\mvec{\kappa}(\Theta,\mvec{R}\cdot\nabla\Theta)\cdot\mvec{R}.
\ee
For a rigid, isotropic and linear heat conductor, this results in
\bee{38}
\mvec{\kappa}(\Theta)\ =\ \kappa(\Theta)\mvec{1}.
\ee

\section{Comment}

The differences between the present paper and that of I-Shih Liu \C{1}
are discussed in catchwords:
\begin{itemize}
\item Rigidity\newline
The domain of the material mapping, the state space, should not contain 
variables of no influence on the material mapping's range, the
constitutive properties. Consequently, neither the deformation
gradient, its rotational part, nor the Cauchy-Green tensors appear
in the state space of rigid materials. Because all these 
quantities are absent,
the material in consideration is rigid. This more ``simple-minded
formulation'' \C{1} seems more directly than introducing variables
having no influence on the material properties, especially if these
variables have be removed later on. The axiomatic basis of the theory
is not weakened by ignoring the above mentioned variables in the state 
space.\vspace{.3cm}
\item Symmetry\newline
In the present paper, the symmetry is formulated in the current
configuration. Its definition is identical to that in the material
configuration. Symmetry is observer-independent, and therefore
described by a tensorial equation.\vspace{.3cm}
\item Material Frame Indifference\newline
Material frame indifference is strictly connected to changing the
observers. Consequently, it is given by a component equation,
where the components belong the observers. Basis of this procedure is
the observer-independence of the tensorial material mapping.
Using erroneously a tensor formulation for material frame indifference induces
isotropy and connects changing of the observers to isotropic materials,
a fact which should be avoided.\vspace{.3cm}
\item Isotropy\newline
Isotropy is an observer-independent symmetry, and consequently also
independent of changing the observers. In \C{1} (and almost
everywhere) isotropy is generated by changing the observer, thus
mixing two items which are axiomatically independent of each other. 
\end{itemize}

\end{document}